# Why are co-authored academic articles more cited: Higher quality or larger audience?

Mike Thelwall
Statistical Cybermetrics and Research Evaluation Group, University of Wolverhampton, UK.
https://orcid.org/0000-0001-6065-205X m.thelwall@wlv.ac.uk

Kayvan Kousha
Statistical Cybermetrics and Research Evaluation Group, University of Wolverhampton, UK.
https://orcid.org/0000-0003-4827-971X k.kousha@wlv.ac.uk

Mahshid Abdoli
Statistical Cybermetrics and Research Evaluation Group, University of Wolverhampton, UK.
https://orcid.org/0000-0001-9251-5391 m.abdoli@wlv.ac.uk

Emma Stuart
Statistical Cybermetrics and Research Evaluation Group, University of Wolverhampton, UK.
https://orcid.org/0000-0003-4807-7659 emma.stuart@wlv.ac.uk

Meiko Makita
Statistical Cybermetrics and Research Evaluation Group, University of Wolverhampton, UK.
https://orcid.org/0000-0002-2284-0161 meikomakita@wlv.ac.uk

Paul Wilson
Statistical Cybermetrics and Research Evaluation Group, University of Wolverhampton, UK.
https://orcid.org/0000-0002-1265-543X pauljwilson@wlv.ac.uk

Jonathan Levitt
Statistical Cybermetrics and Research Evaluation Group, University of Wolverhampton, UK.
https://orcid.org/0000-0002-4386-3813 j.m.levitt@wlv.ac.uk

Co-authored articles tend to be more cited in many academic fields, but is this because they tend to be higher quality or is it an audience effect: increased awareness through multiple author networks? We address this unknown with the largest investigation yet into whether author numbers associate with research quality, using expert peer quality judgements for 122,331 non-review journal articles submitted by UK academics for the 2014-20 national assessment process. Spearman correlations between the number of authors and the quality scores show moderately strong positive associations (0.2-0.4) in the health, life, and physical sciences, but weak or no positive associations in engineering, and social sciences. In contrast, we found little or no association in the arts and humanities, and a possible negative association for decision sciences. This gives reasonably conclusive evidence that greater numbers of authors associates with higher quality journal articles in the majority of academia outside the arts and humanities, at least for the UK. Positive associations between team size and citation counts in areas with little association between team size and quality also show that audience effects or other non-quality factors account for the higher citation rates of co-authored articles in some fields.

# Introduction

A key factor behind much science policy is encouragement for increased collaboration (Lee & Bozeman, 2005; Ubfal & Maffioli, 2011). For example, European Union research funding is often predicated on the participation of at least three different countries, Marie Curie and Fulbright scholarships fund mobility, and large-scale instruments require extensive funding and cooperation (D'ippolito & Rüling, 2019). This policy is underpinned by historically influential theory arguing that big science or interdisciplinary collaboration are necessary to address complex societal challenges (e.g., de Solla Price, 1963; Gibbons et al., 1994; Ziman, 2000), and the importance of academic teamwork is recognised by a dedicated research field about team science (Hall et al., 2018).

The value of collaboration is supported by a substantial body of empirical research showing that articles tend to be more cited when they have more authors (e.g., most studies listed in: Shen et al., 2021), although the science-wide citation advantage of co-authorship decreased 1900-2011 (Larivière et al., 2015). Nevertheless, it has been argued co-authored articles tend to be more cited mainly because the additional authors attract more interest through their personal networks, particularly when multiple countries are represented (Deichmann et al., 2020; Ding et al., 2021; Wagner et al., 2019), rather than co-authored studies tending to be of better quality. Moreover, solo authorship is valued in the arts and humanities (e.g., Hansson et al., 2021) so it is unclear whether co-authorship would have a similar association with quality in all fields. As reviewed below, two previous empirical analyses of the relationship between co-authorship and research quality (Bornmann, 2017; Franceschet & Costantini, 2010) have found weak relationships but larger studies are needed to give detailed information about disciplinary differences.

This article reports the largest empirical analysis yet of the relationship between quality and the number of co-authors of academic research, filling the gaps identified above. It follows a previous similar study (Franceschet & Costantini, 2010) with a different country (UK rather than Italy), six times more articles, 15 years more recent publications, finer grained and multiple field classifications (34, 27 and 22 fields rather than 10), and with evidence of statistical associations that is comparable between fields (correlations rather than group-based chi square tests). It addresses the following research questions.

- RQ1: In which fields does collaboration associate with higher research quality?
- RQ2: What is the relationship between the number of authors and research quality?
- RQ3: In which fields does collaboration associate with more citations but not higher research quality?

# Background

Co-authorship tends to be used to acknowledge involvement in a study, although the nature and extent of contributions varies between fields (Larivière et al., 2016), and there are exceptions like ghost and gift authorship (Gülen et al., 2020). Collaboration does not necessarily lead to co-authorship of a journal article, but the two are conflated here for simplicity. The reasons why researchers may decide to collaborate include social pleasure, PhD supervision, junior research visits, funding, and access to equipment (Katz & Martin,

1997; Melin, 2000; Shen et al., 2017; Sonnenwald, 2007; Ubfal & Maffioli, 2011). Researchers may also collaborate to learn from each other or to provide complementary skills (e.g., statistics and survey design). Co-authorship may be effectively mandatory, for research funding or PhD supervision, or optional, as when two mathematicians take turns to tackle a problem of shared interest. Thus, co-authorship covers a range of different phenomena. The question of whether collaboration is advantageous overall is therefore asking whether the contexts in which it is either advantageous or necessary for high quality research are more prevalent than the contexts in which it is disadvantageous or necessary for low quality research. On this basis, there can be no overall theory of collaboration advantages.

## *Co-authorship and citation counts*

Positive associations between author numbers and citation-based indicators have been found in many contexts, always grouping all types of co-authorship together, and this section briefly reviews relevant large-scale studies. In general, co-authorship has been found to associate with higher citation counts in most contexts analysed, but with exceptions covering whole countries, whole disciplines, and country/discipline combinations. The exceptions could be differing combinations of types of collaboration. For example, the results might be weaker in countries/fields/eras when collaboration was not incentivised with funding, was primarily domestic, or was primarily for PhD research.

For Web of Science articles 2000-2009 in Biology and Biochemistry, Chemistry, and Social Sciences, there is a positive correlation between author numbers and citation counts (Didegah & Thelwall, 2013). For Italian-authored research 2004-2010, linear regressions revealed a positive association between the number of authors and article citations (percentiles within Italy) for all areas examined: Biology; Biomedical research; Chemistry; Clinical medicine; Earth and space sciences; Economics; Engineering; Law, political and social sciences; Mathematics; Multidisciplinary sciences; Physics; Psychology (Abramo & D'Angelo, 2015). The association was strongest in Clinical Medicine and weakest in Engineering. For South African papers from 2000, 2003 and 2005, collaborative articles were more cited than solo articles in most areas of academia, but the opposite was true in Psychiatry, Biochemistry, Agriculture, and Material Science (Sooryamoorthy, 2009). The latter fields are not exceptions in other studies. For the nine countries with the most articles in Scopus, domestic collaborative research (all authors from the same country) was more cited than solo research except in Russia (Thelwall & Sud, 2016). The same study found that collaboration was associated with more citations in arts and humanities much more strongly than in business, chemistry, and pharmaceutics. A larger-scale follow-up study of ten countries (not including Russia) found the weakest association between citations and author numbers to be in China (Thelwall & Maflahi, 2020). It also found a general trend for domestic research to be more cited when it had multiple authors, although there were exceptions, such as business research in Germany. International examples of situations without an association between citations and collaboration include finance 1987–1991 (Avkiran, 1997). Thus, whilst greater numbers of authors associates with higher citation rates in most countries and fields, there are exceptions for unknown reasons.

There are many possible causes of collaborative research tending to be more cited (Hall et al., 2018). The greater expertise brought by a team of people might allow their work to be higher quality overall – for example because each is a specialist for their task - or they may be able to tackle more important problems with larger or more interdisciplinary teams (Gibbons et al., 1994). At a statistical level, the presence of a few articles from an influential large, highly-funded team (Thelwall, 2020) can induce a positive correlation in a set of articles that otherwise would not show a trend. Alternatively, each author may bring an audience of people that know them or that are interested in their work, so larger teams may generate more readers. This may be especially true for international co-authorship reaching different countries (Lancho Barrantes, et al., 2012; Schmoch & Schubert, 2008; Wagner et al., 2019).

*Co-authorship and research quality*

Research quality is judged during peer review before publication, national research evaluations, post publication informal evaluations by readers, and job applications. Peer review has become the standard method of assessing the quality of journal articles, particularly prior to publication (Benos et al., 2007). There are three generally agreed main components of research quality: rigour, novelty/originality, and significance to science or society (Langfeldt et al., 2020). These dimensions vary in how they are assessed between fields (e.g., REF, 2019), despite being relevant to all.

There are many reasons why articles with a greater number of authors tend to be better in some aspect of quality, at least in some disciplines, and some reasons that point in the opposite direction (Beaver, 2013; Hall et al., 2018). Most obviously, collaboration increases the pool of expertise and possibly also resources for an article but may reduce the strength of the weakest link in a team or reduce the chance of radical discoveries by maverick scientists (Hull, 2010). There is no single cause of the collaboration citation advantage because of the variety of reasons for collaboration, disciplinary differences in research types, varied incentives to collaborate, and multiple benefits from collaboration (Van Rijnsoever & Hessels, 2011).

In some biological (Cantor et al., 2010) and medical research (e.g., vaccine trials), increasing the sample size for human subjects research is expensive and complex because it requires the work of more people to be coordinated. Thus, larger-scale studies are both harder to achieve and more likely to generate positive results. Other factors being equal, increased research quality (particularly in terms of significance), can be expected with larger numbers of co-authors in this case. In contrast, some arts and humanities work might primarily require the deep and creative thought of an individual scholar, who may be sceptical about the value of collaboration (e.g., as mentioned in: Real, 2012; the importance of [usually solo] monographs: Shaw et al., 2022). Similarly, in some cases a collaboration might indicate that a senior scholar had delegated some of the work, such as the literature review or one experiment, to a junior colleague, perhaps affecting its quality.

From a different perspective, some research goals, such as vaccine development (Gilbert & Green, 2021) and high energy physics experiments (Perkins, 2000), are fundamentally big science endeavours that require large scale coordination of different specialisms. These are impossible to conduct on a small scale, so if their outputs are judged to be high quality – for example as relatively unique contributions to science - then this alone would generate a statistical association between team size and research quality.

Funding is important in many research fields, and a better funded study might operate on a larger scale, producing higher quality, more useful results. If part of the funding is spent on creating a larger team, then this would create an association between team size and quality. Funding in medicine associates with higher rates of citation, partly because of the larger team sizes of funded research (Yan, Wu, & Song, 2018).

From yet another perspective, interdisciplinary research must include a relatively unusual combination of input methods or expertise (e.g., Wagner et al, 2019) and therefore seems likely to be judged to have higher methodological originality and may also be more likely to generate societal impacts (e.g., Gibbons et al., 1994; Ignaciuk et al., 2012).

Figure 1 illustrates the complexity of the relationship between collaboration and research quality, summarising some of the above points. Whilst more authors would bring more expertise and resources, which should increase research quality, they may also weaken the weakest link, increase conservatism, and bring coordination problems, which may detract from research quality. On the other hand, expensive equipment and large funding pots might generate large teams and good research (because better resourced). Also, a better researcher might produce better research and attract more co-authors. In any given field there may also be a tendency for big team types of research to be high or low quality. For example, Hadron Collider studies, with typically thousands of authors, might be generally regarded by high energy physicists as usually excellent or usually routine. For PhD supervision, the research might tend to be higher or lower quality in some fields, and the PhD student may tend to work in fields that are larger or smaller than the field average, so this relationship may well vary between fields.

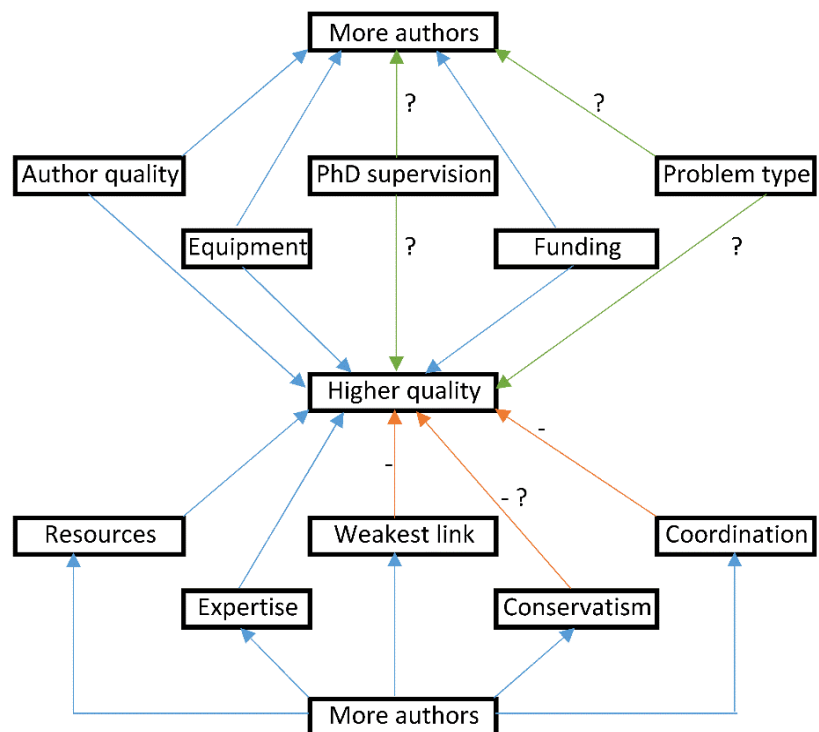

Figure 1. Some factors that might directly or indirectly contribute to the relationship between collaboration and research quality in a field. Blue arrows indicate a positive relationship, orange negative, and green unknown or variable between fields.

*Empirical evidence of relationships between co-authorship and research quality*

Given the complex multiple factors at work, empirical evidence is needed to identify the fields in which there is a clear association between collaboration and research quality. There is limited indirect evidence of this from a few small scale studies. One old study found that articles with more authors were less likely to be rejected during peer review in a psychology journal (Presser, 1980). Graduate library and information science students tended to find collaborative journal articles more useful for their course than solo authored journal articles (Finlay et al., 2012), addressing the significance dimension of quality. If journal impact is an acceptable proxy for research quality, then New Zealand biomedical scientists tend to produce higher quality research when they collaborate (He et al., 2009).

Two studies have investigated the relationship between author numbers and peer reviewed journal article quality on a larger scale. An analysis of 16,554 biology and medicine articles from before 2014 found that the (field and year normalised) citation advantage of collaboration could not be explained by the quality of collaborative articles, as judged by F1000Prime (now Faculty Opinions) reviewers. There was a very weak Spearman correlation (0.09) between article quality and the number of authors per paper (Bornmann, 2017). A large-scale descriptive multidisciplinary study of 18,500 Italian research outputs 2001-2003 rated by expert peer review found that articles with more authors tended to have higher quality ratings in most fields: Physics; Earth Sciences; Biology, Medical Sciences; Agricultural Sciences and Veterinary Medicine; Philological-literary Sciences, Antiquities and Arts; History, Philosophy, Psychology, and Pedagogy; Economics and Statistics; and Political and Social Sciences. There was not a clear positive relationship in a few fields: Chemistry; Mathematics and Computer Sciences; Civil Engineering and Architecture; Industrial and Information Engineering; and Law (Franceschet & Costantini, 2010). Similar authorship team sizes were grouped together for this analysis, which may have hidden finer-grained relationships, and six of the differences were statistically significant. Thus, larger scale studies with multidisciplinary data and covering other countries are needed to give finer grained information and a different national perspective.

## Methods

### Data

We obtained provisional quality scores for 148,977 journal articles submitted to the UK REF2021 from the REF team under a confidentiality agreement as part of a research contract in March 2022, excluding submissions from the University of Wolverhampton for privacy reasons. REF journal articles must be primary research rather than reviews and are the self-selected best 1-5 outputs of all active UK researchers first published between 2014 and 2020. We were not given information about the other output types submitted by researchers (e.g., books, software, performances). The data includes scores on a four-point scale for overall quality: 1*, 2*, 3*, or 4*. Articles were scored 0 if they did not qualify for review or were judged not to be research. The scores are allocated by sub-panels of subject experts, usually with at least two per output, who promise to read each output before making their decision. Sub-panels meet for quality control purposes and evaluators are given clear guidelines about how to evaluate the quality of the research in their UoA.

We discarded all 318 articles with score 0 since at least some were high quality articles that had been judged out of scope for authorship reasons or type reasons (judged to be a review). The remaining articles included many duplicates, due to articles submitted by multiple authors. We eliminated these duplicates separately within each analysis unit (UoA or main panel), retaining the median score when an article had been given different scores within the unit analysed (chosen at random when there were two medians).

We matched the REF articles with Scopus articles published 2014-20 primarily by DOI but also by title and authors for additional matching (similar to: Bornmann, 2017), with human checks of these additional results, giving 997 new matches. We discarded non-matching REF articles. We used the Scopus matches to record the number of authors for each article, as listed in Scopus, and for the Scopus broad fields of each article. Scopus assigns articles to broad fields usually based on the publishing journal. When a journal was assigned to multiple fields then all its articles were also assigned to all those fields. This produced 122,331 REF papers scoring 1* to 4* and matching a Scopus record 2014-20.

We also searched for the 122,331 articles that matched with Scopus in the scholarly search engine Dimensions.ai (Hook et al., 2018) through DOIs, although it was not practical to also check by title for the missing articles. All articles for Dimensions.ai also had to be in Scopus to use the Scopus author counts, which seemed to be the most reliable. This restriction is not a major problem because Dimensions indexes 84% of Scopus (Martín-Martín et al., 2021). We used the Dimensions records only for its top-level Field Of Research (FOR) codes, which are assigned to articles on an individual basis (Dimensions.ai, 2021), with each article typically assigned multiple broad and/or narrow codes. We excluded articles that had not been assigned a FOR code (because the Dimension field classification AI software had not reached its probability threshold).

After these processing steps, there were 134,801 articles in Dimensions broad fields, multiply counting articles with more than one broad FOR code, 122,331 articles in UoAs, multiply counting articles submitted by authors to more than one UoAs, and 201,635 articles in Scopus, multiply counting articles in more than one Scopus broad field.

For the second research question, raw Scopus citation counts were available for all the articles analysed but these would not be appropriate because (a) the data is from multiple years and (b) UoAs combine multiple fields. Thus, we transformed the citation counts into field normalised scores, the Normalised Log-transformed Citation Score (NLCS) (Thelwall, 2017). For this, we first log transformed all citation counts with ln(1-x) to reduce skewing and support the calculation of more precise averages. Next, we calculated the average log-transformed citation count for every Scopus narrow field and year (326 fields x 7 years = 2,282 averages, with some empty field/year combinations), based on all standard journal articles, not just the REF2021 articles. We then calculated the NLCS for each article as its log-transformed citation count ln(1+x) divided by the average for its field and year. Articles in multiple fields were instead divided by the average of the averages for each of the fields. The NLCS for each article is 1 if the article had the average number of citations for all articles in its field(s) and years. Scores greater than one indicate above average citation impact and scores below one indicate below average citation impact, irrespective of field(s) and year. By design, is fair to compare NLCSs for articles in different fields and years since each NLCS calculation is norm referenced only against its field(s) and year.

## *Analysis*

We assessed the relationship between the number of authors and the quality of an article (1* to 4*) using Spearman correlations. This approach was used instead of regression to investigate the overall pattern rather than the contribution of coauthorship relative to other factors or the effect of different types of collaboration. We log transformed author numbers with ln(1+x) and Pearson correlations were also used with the log-transformed data. Pearson correlations are more problematic because the four-point REF scoring system is an ordered set of qualitative categories rather than a scale. The results of the two types of correlation were similar so we only report the Spearman correlations for simplicity. 95% confidence intervals were calculated using the Fisher (1915) transformation.

We calculated the same correlations for author numbers and citation counts. The two sets of Spearman correlations are not directly comparable because of the large difference in granularity between NLCS and quality scores, which affects the magnitude of a correlation for a similar underlying relationship strength (Thelwall, 2016).

## Results

### *RQ1: Author counts and REF2021 quality scores*

At the level of the Units of Assessment (UoAs) in which research outputs were evaluated in REF2021, there are statistically significant moderate positive correlations (0.2-0.4) between author counts and research quality in all health, life, and physical sciences (Figure 2). There are also moderate correlations (above 0.2) in two social sciences: Archaeology, and Sport, Exercise Sciences, Leisure, and Tourism (possibly party due to the inclusion of medical topics within sports research). The confidence intervals tend to be quite wide for the other fields, but most social science, arts and humanities areas have weak or possibly no correlation between co-authorship and research quality. The data is consistent with some humanities areas having a weak negative association between co-authorship and research quality, but the confidence intervals for the negative correlations all include 0.

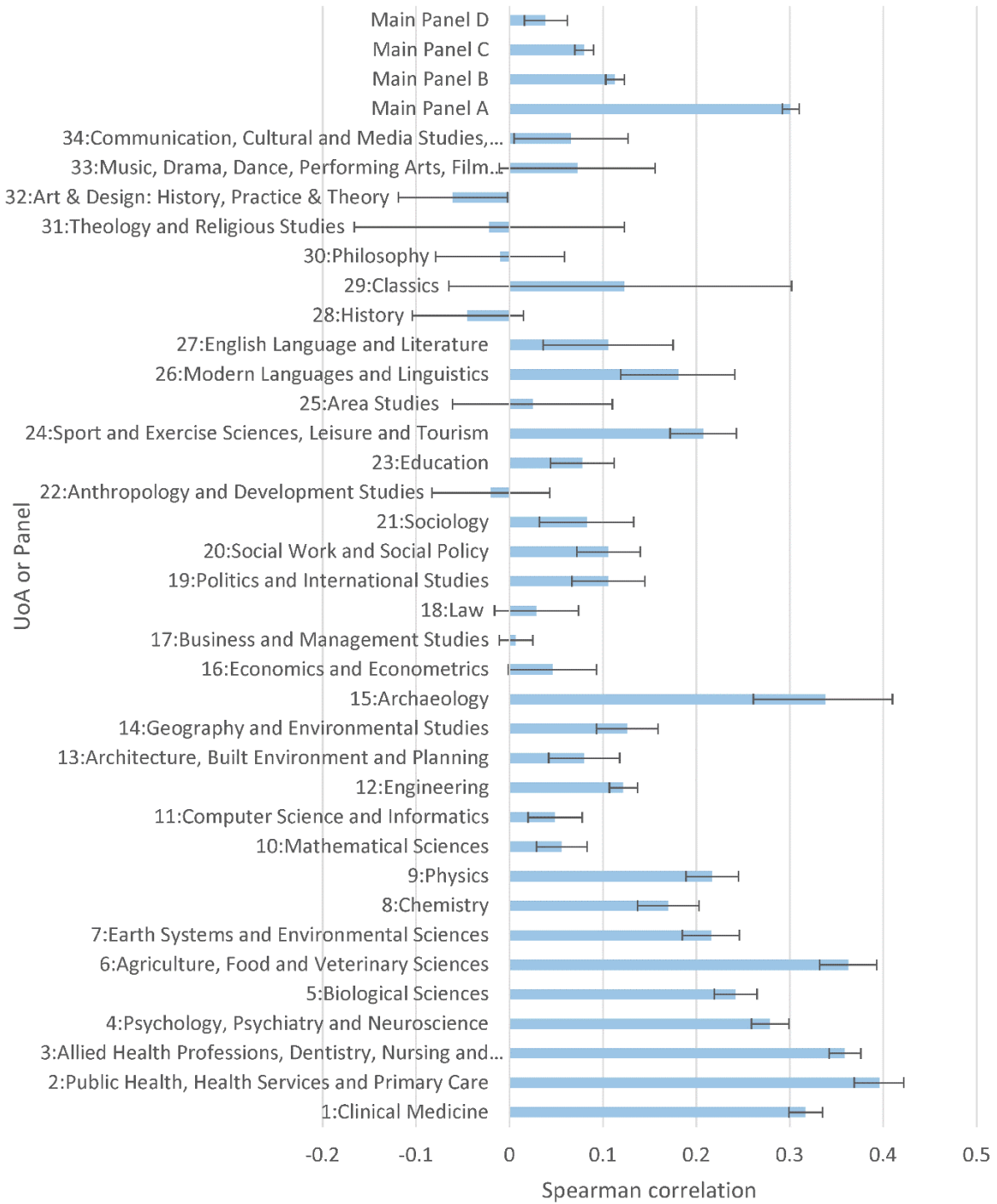

Figure 2. Spearman correlation between REF quality score and number of authors, as recorded in Scopus, for REF2021 articles published 2014-20, by UoA or Main Panel. Main panels include all articles in their UoAs, after eliminating duplicates.

When Scopus broad fields are used for the correlations instead of REF UoAs, the pattern is similar (Figure 3) but there is evidence of a negative association between co-authorship and research quality in Decision Sciences. This broad field includes three quite diverse subfields: Information Systems and Management; Management Science and Operations Research; and Statistics, Probability and Uncertainty. It is possible that the negative association is due to one of these fields tending to score higher in the REF (e.g., due to one large strong department) and tending not to co-author. The correlation is statistically significantly positive for the Arts and Humanities, but this is not strong evidence because the set may include articles from other fields that are additionally classified as Arts and Humanities (e.g., electronic music technology research).

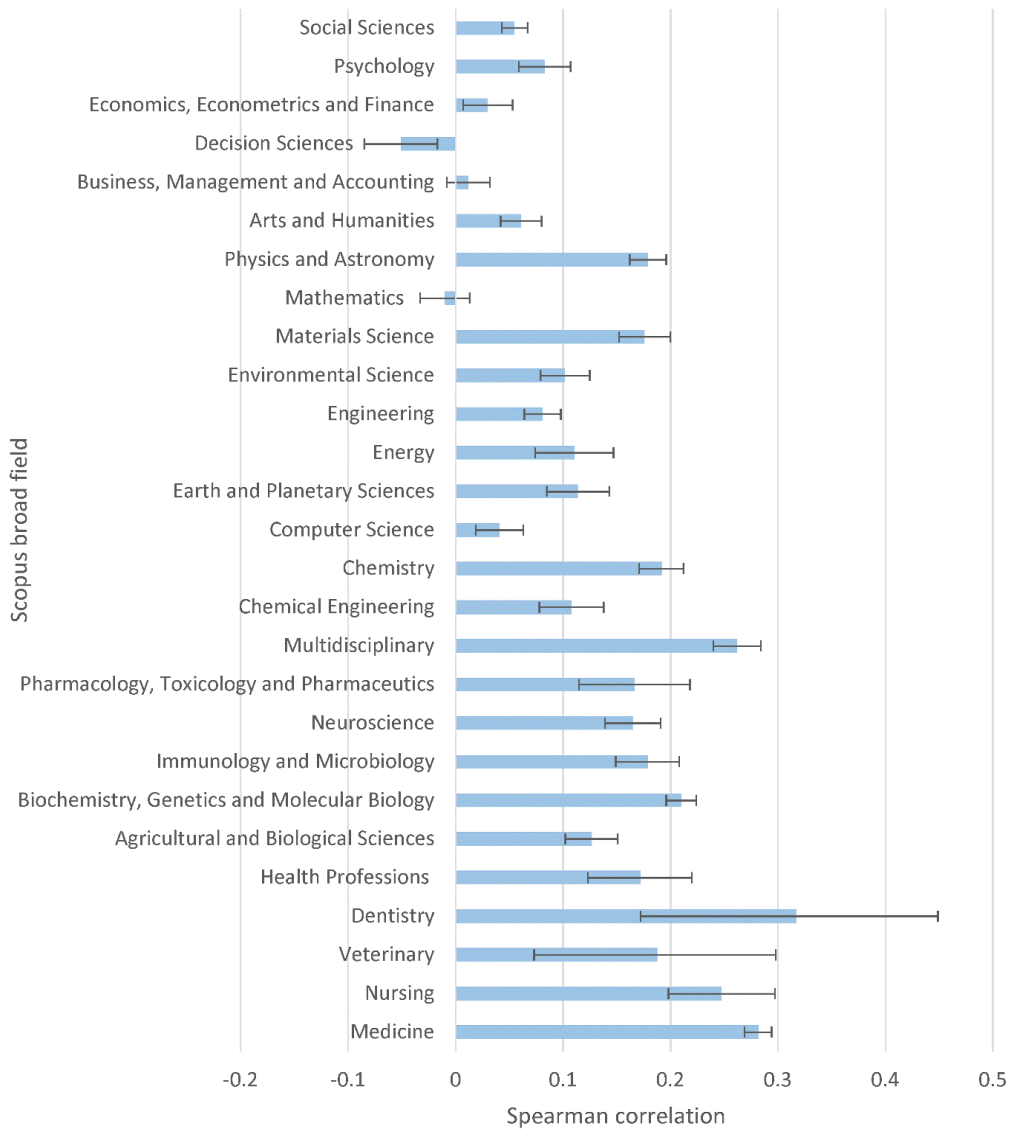

Figure 3. Spearman correlation between REF quality score and number of authors, as recorded in Scopus, for REF2021 articles published 2014-20, by Scopus broad field (grouped by top level field).

When using the Dimensions FOR codes, the results are again similar to before (Figure 4). All correlations are either positive or include zero in their confidence intervals. The arts are not represented, with the partial exception of creative arts and perhaps aspects of culture, and two of the three predominantly humanities categories (19 and 22, but not 18) have negative correlations.

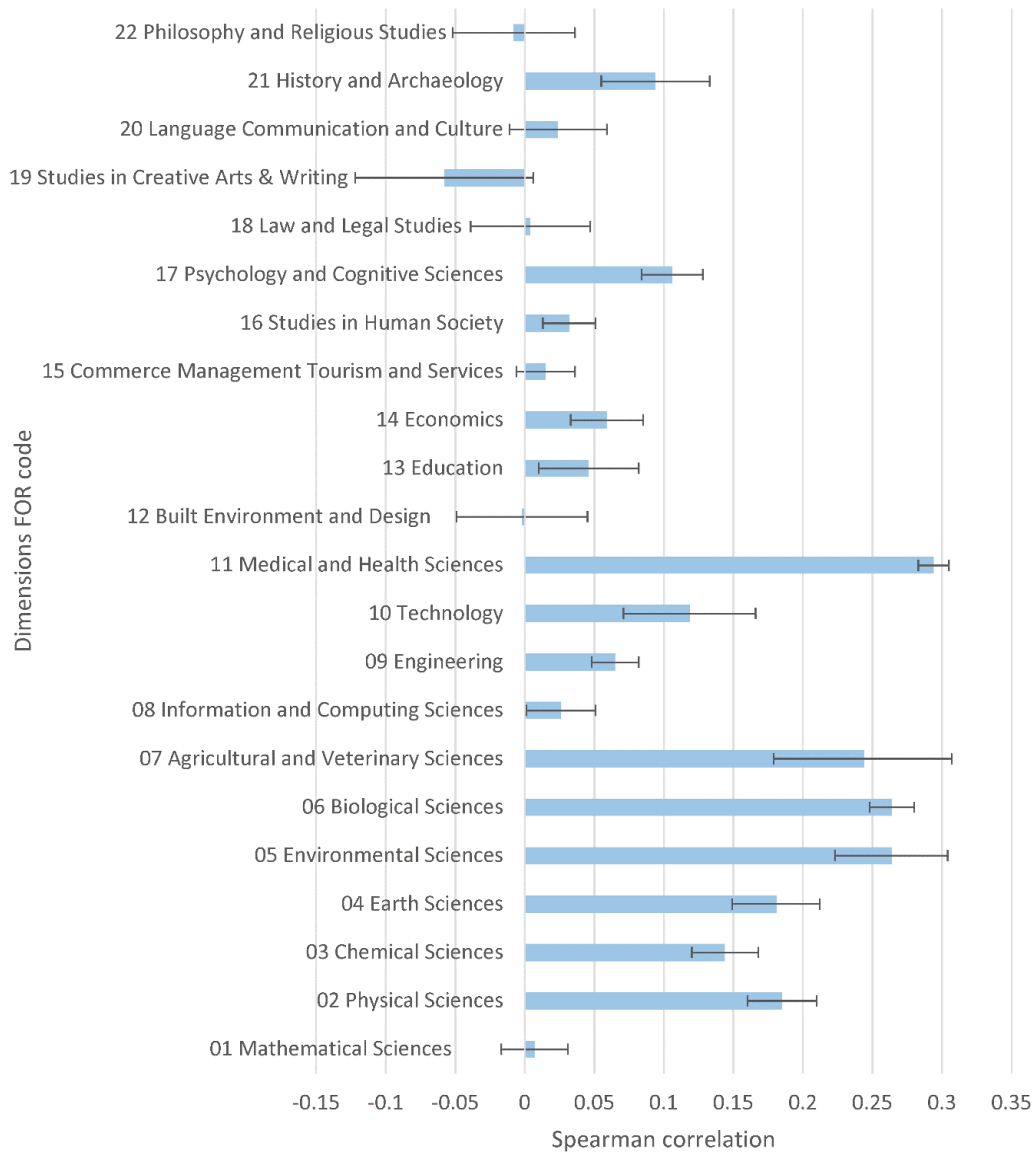

Figure 4. Spearman correlation between REF quality score and number of authors, as recorded in Scopus, for REF2021 articles published 2014-20, by Dimensions FOR code.

## RQ2: Author numbers and research quality

The underlying shape of the relationship between the number of authors and the REF quality score of articles is approximately logarithmic for fields with a positive correlation, as illustrated by UoA 1 and UoA 12 (Figure 5). Although the correlation is weak for UoA 12, the trend is strong, at least for low numbers of authors. The reason for the low correlation for UoA 12 is the wide range of different REF scores for low numbers of authors. The shape for UoA 32 suggests that single author research is valued, as is common in the humanities.

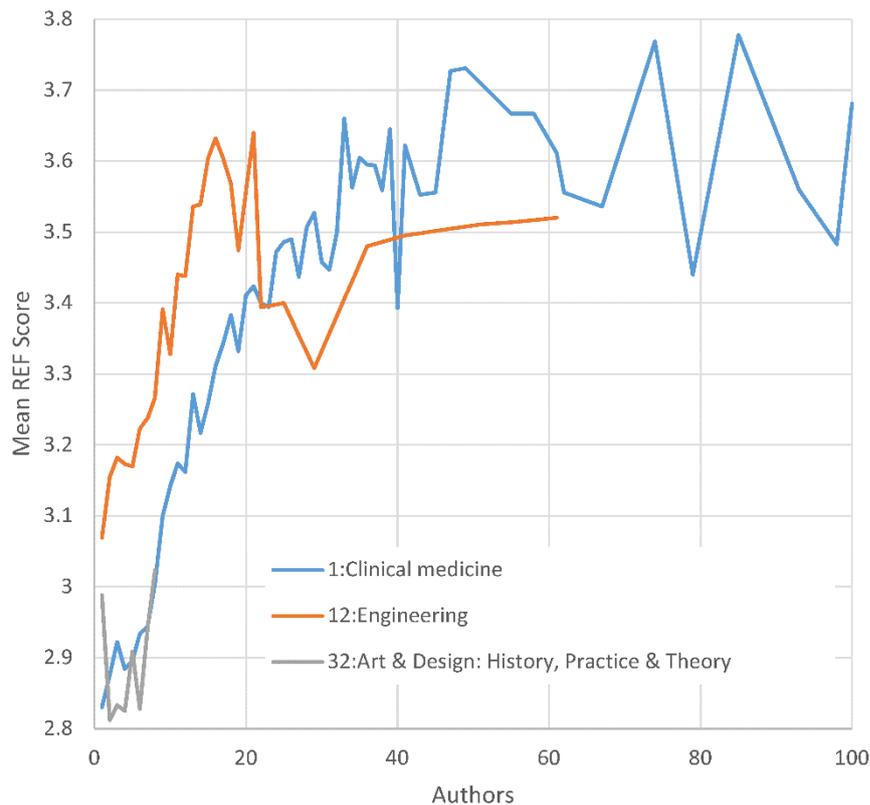

Figure 5. Average REF quality score against number of authors, as recorded in Scopus, for REF2021 articles published 2014-20, for three illustrative UoAs.

### RQ3: Author counts and field normalised citation counts

In contrast to the situation for REF quality scores, for the same set of articles there is a statistically significant positive correlation between the number of authors and the field normalised citation score for *all* UoAs (Figure 6). For one UoA (32) the 95% confidence interval in Figure 2 excludes positive values and the 95% confidence interval in Figure 6 is exclusively positive. Thus, for this UoA it is possible to claim that increased co-authorship does not associate with increased quality but does associate with increased citations. The fact that the correlations are statistically significantly positive in all UoAs in Figure 6 and are sometimes negative in Figure 2 and are often close to zero also gives additional collective evidence that it is possible for increased citations to associate with more authors irrespective of article quality (i.e., possible audience effects). The situation is confirmed for Scopus broad fields (Figure 7), except in the case of Veterinary due to its small sample size.

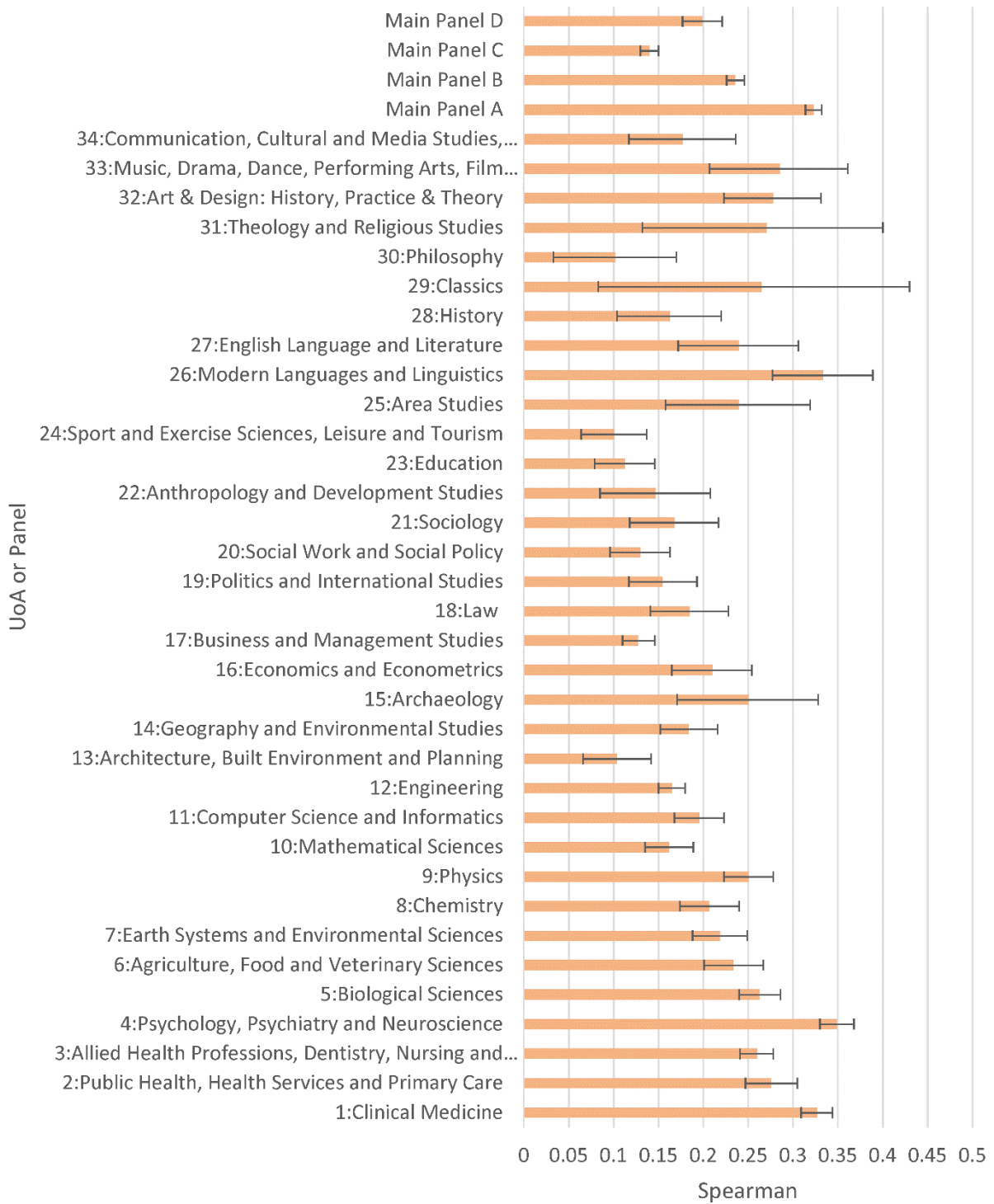

Figure 6. Spearman correlation between field normalised Scopus citation count and number of authors, as recorded in Scopus, for REF2021 articles published 2014-20, by UoA or Main Panel. Main panels include all articles in their UoAs, after eliminating duplicates.

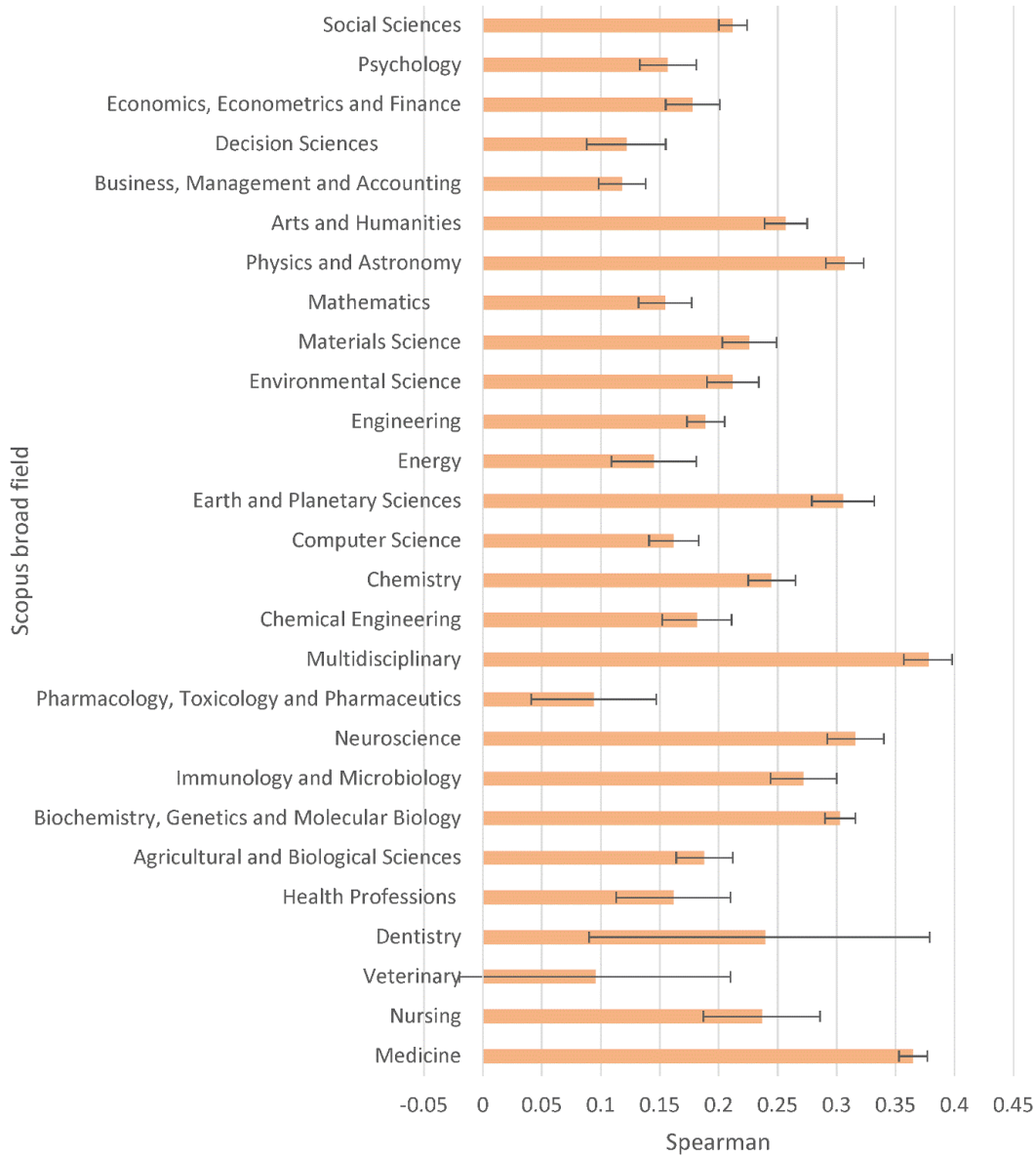

Figure 7. Spearman correlation between field normalised Scopus citation count and number of authors, as recorded in Scopus, for REF2021 articles published 2014-20, by Scopus broad field (grouped by top level field).

With the Dimensions FOR codes, there are some areas of the humanities (18, 19, 22, some of 20, 21) where collaboration has quite a strong association with Scopus citation counts, even though there is little or no association between collaboration and research quality (Figure 8). Perhaps more collaborative research tends to be in more highly cited areas, such as those with elements of the social sciences or other sciences.

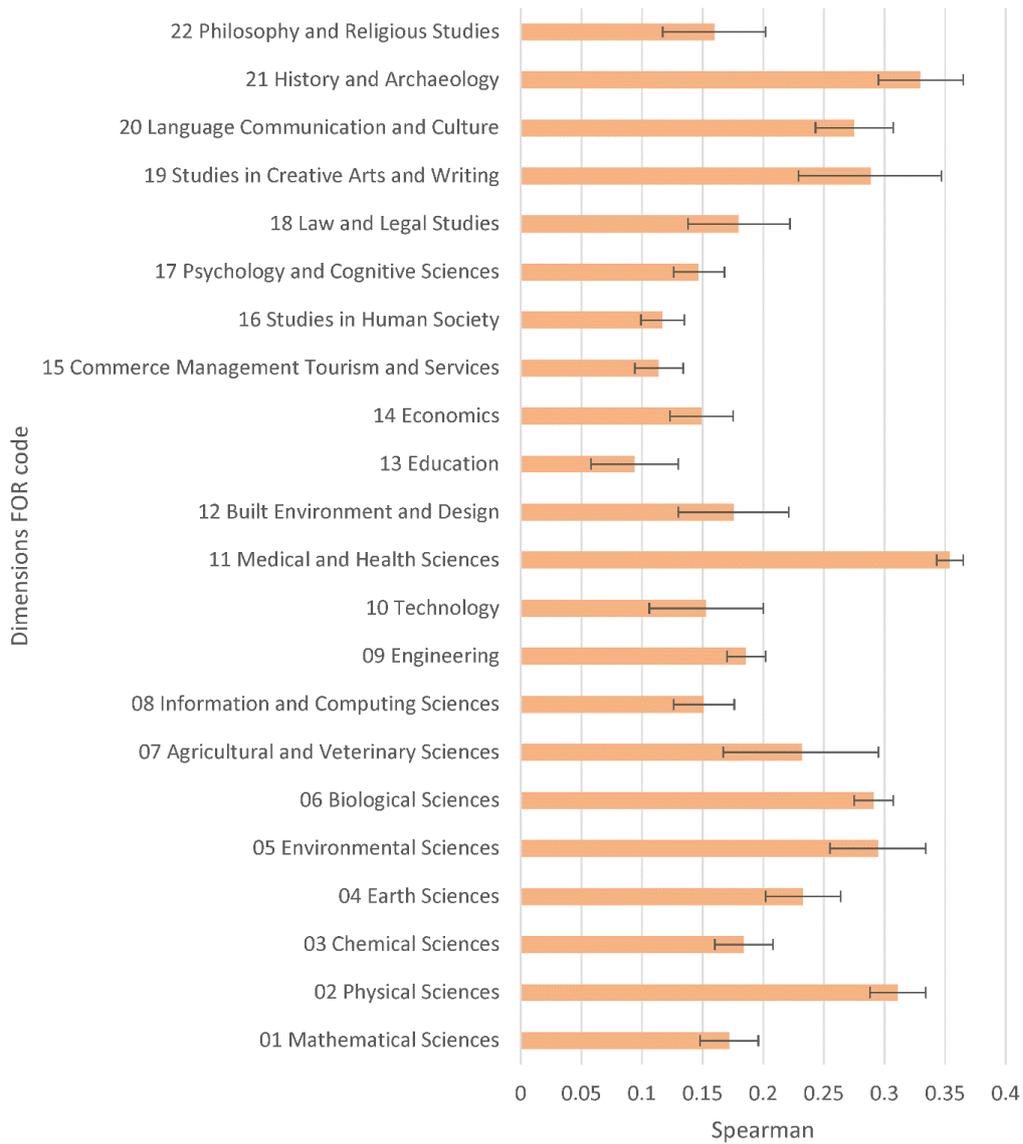

Figure 8. Spearman correlation between field normalised Scopus citation count and number of authors, as recorded in Scopus, for REF2021 articles published 2014-20, by Dimensions FOR code.

The underlying shape of the relationship between the number of authors and field normalised citation counts NLCS tends to be neither approximately linear nor approximately logarithmic (e.g., Figure 9), in contrast to the situation for quality scores. The slight difference in shapes suggests that factors other than quality influence the relationship between citations and author numbers.

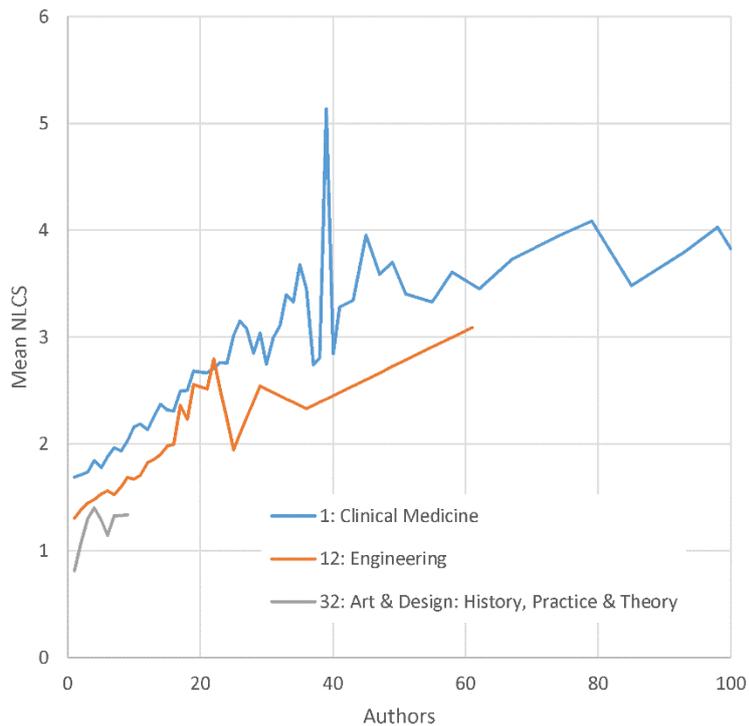

Figure 9. Average field and year normalised citation score (NLCS) against number of authors, as recorded in Scopus, for REF2021 articles published 2014-20, for three illustrative UoAs.

## Discussion

The results are limited by the UK focus. Although international articles are included, at least one author must have had a UK affiliation by July 2020. Co-authorship has international differences in relationships with citations (Thelwall & Maflahi, 2020), so it is possible that different relationships between quality and author numbers might be found in some countries or for domestic research. The data is also self-selected to be the highest quality outputs of UK academics, which probably reduces the strength of the correlations found, due to relatively few articles having low scores. Similarly, the relatively coarse-grained quality scheme limits the practical possibility of obtaining high correlations. The results also say nothing about review articles, books, and other research outputs.

Our results disagree with a previous study using a different method (Bornmann, 2017) and partly agree with the similar prior study (Franceschet & Costantini, 2010), in terms of approximately matching UoAs to the Italian field groupings, but our results show a clear positive relationship between author numbers and research quality in areas related to the VQR categories of Chemistry, Mathematics and Computer Science (although very weak), Civil Engineering and Architecture, and Industrial and Information Engineering. Nevertheless, there was little or no relationship between author numbers and research quality for Law in both the UK and Italy.

For fields in which the above results suggest that there is little association between team size and research quality, but other research demonstrates a strong relationship between team size and citation counts, it would be reasonable to conclude that the extra citations are caused by audience effects of team size (Wagner et al., 2019) rather than increased research quality.

The correlation results do not prove a cause-and-effect relationship in the sense that adding extra members to a team for the same research would tend to improve its quality.

There are many alternative explanations, as summarised in the top half of Figure 1. For example, team leaders with a track record of high quality research may be funded to attract more collaborators. The diversity of types of research within all the categories reported above (e.g., theoretical vs empirical; different topics; different methods), make it impossible to find a clear explanation for the relationships found.

Cause-and-effect relationships are also plausible, as summarised in the bottom half of Figure 1. For example, in empirical areas in the health domain, larger numbers of co-authors may reflect either larger scale studies (more data collection centres) or more complex analyses (more methods), which would tend to produce more powerful findings. In contrast, in arts and humanities and some social science topics, teams may often represent mentor/mentee relationships or groups of colleagues with a common interest and similar expertise topics rather than tending to enable more powerful analyses. Since funding encourages collaboration (Davies et al., 2022), successful researchers in areas with less project-based funding may also prefer to work alone and may not need teams of people to help their research process.

The evidence of moderate positive correlations between author numbers and citations in some fields for which there are little, no, or negative correlations between author numbers and article quality shows that non-quality factors must be the cause of increased citations for larger team papers in some fields. This is supported by the slightly different shapes in the relationships between author numbers and citations or quality scores, even in fields for which both correlations are positive. The previously hypothesised audience effects are plausible, and the results suggest that they are more important than quality in some fields and the reverse – or possibly irrelevant - in others. Nevertheless, the results may also be due to non-audience, non-quality factors, such as a possible tendency for authors to collaborate more on more citeable topics.

## Conclusion

The results show, for the first time, the strength of association between team size and research quality across all areas of science for non-review journal articles, albeit with a UK focus. They give clear evidence of larger teams associating moderately with higher quality research, as judged by careful consideration from REF field experts, in the health, life, and physical sciences. They show that weak positive associations are present for engineering, and many social sciences, but that there is little or no association in the arts and humanities. There is also evidence of a negative association in decision sciences.

The results also show that factors other than article quality are needed to explain the tendency for articles with more authors to be more cited in some fields. A possible explanation is that more authors attract greater attention to research in some fields.

The results do not prove that all types of collaboration are beneficial or reveal which types are beneficial. Thus, they are not useful for individual decisions about whether to add extra scholars to a team or whether it would be beneficial to create a larger group for a study. They also do not address the issue of research productivity. Nevertheless, they are relevant for large scale policy decisions that encourage collaboration on the basis that, in general, it is beneficial to scientific quality, and not just citation impact.

Although the results do not prove cause-and-effect, they tend to confirm the importance of collaboration in the health, life, and physical sciences, which should reassure funders that mandate team formation or that encourage larger consortium bids. They also suggest that in engineering, the social sciences, and the arts and humanities, a more cautious approach should be taken because high quality research is almost as likely to be conducted by smaller and larger groups of co-authors. It would be a shame to provide strong collaboration incentive to researchers when there is little benefit because this might reduce research diversity and restrict the opportunities for researchers to investigate topics that do not need extensive teamwork.

## Acknowledgement
This study was funded by Research England, Scottish Funding Council, Higher Education Funding Council for Wales, and Department for the Economy, Northern Ireland as part of the Future Research Assessment Programme (https://www.jisc.ac.uk/future-research-assessment-programme). The content is solely the responsibility of the authors and does not necessarily represent the official views of the funders.